\documentclass[aps,prl,twocolumn,superscriptaddress,floatfix]{revtex4-1}
\usepackage{graphicx}
\usepackage{color}
\usepackage{float}
\usepackage{rotating}

\newcommand{\dR}{$\Delta R$/$R$}

\newcommand{\EYMO}{Eu$_{0.55}$Y$_{0.45}$MnO$_{3}$}

\begin{document}

\title{Photo-creating supercooled spiral-spin states in a multiferroic manganite}

\author{Y. M. Sheu}
\affiliation{Department of Electrophysics, National Chiao Tung University, Hsinchu, 300, Taiwan}
\affiliation{RIKEN Center for Emergent Matter Science (CEMS), Wako, Saitama 351-0198, Japan}
\author{N. Ogawa}
\affiliation{RIKEN Center for Emergent Matter Science (CEMS), Wako, Saitama 351-0198, Japan}
\author{Y. Kaneko}
\affiliation{RIKEN Center for Emergent Matter Science (CEMS), Wako, Saitama 351-0198, Japan}
\author{Y. Tokura}
\affiliation{RIKEN Center for Emergent Matter Science (CEMS), Wako, Saitama 351-0198, Japan}
\affiliation{Department of Applied Physics and Quantum-Phase Electronics Center (QPEC), University of Tokyo, Tokyo 113-8656, Japan }
%\date{\today}
\begin{abstract}
%Ferroelectrics of magnetic origin promises to have distinct magnetoelectric effect due to the inherent cross coupling between magnetism and ferroelectricity.
We demonstrate that dynamics of the $ab$-spiral-spin order in a magnetoelectric multiferroic \EYMO~ can be unambiguously probed through optical second harmonic signals, generated via the spin-induced ferroelectric polarization. In the case of relatively weak photoexcitation, the ferroelectric and the spiral-spin order remains interlocked, both relaxing through spin-lattice relaxation in the non-equilibrium state. When the additional optical pulse illuminating sample is intense enough to induced a local phase transition thermally, the system creates a metastable state of the $bc$-spiral-spin order (with the electric polarization $P$//$c$) via supercooling across the first-order phase transition between $ab$- and $bc$-spiral. The supercooled state of $bc$-spiral spin is formed in the thermodynamical ground state of $ab$-spiral ($P$//$a$), displaying a prolonged lifetime with strong dependence on the magnetic field along the $a$-axis. The observed photo-switching between the two distinct multiferroic states sheds light on novel photoinduced phenomena in spiral-spin multiferroics.
\end{abstract}

%\pacs{78.47.jh,75.85.+t,72.25.Rb,77.84.-s}
\maketitle

Spin kinetics via light-matter interaction has been of great interest ever since the discovery of ultrafast demagnetization in a ferromagnetic metal \cite{Beaurepaire1996PRL}.   The exchange interaction, spin-orbit interaction, and spin precession are now believed to dominate spin relaxation after photo-excitation in various systems \cite{Kirilyuk2010Review}. The recent development of spin-induced magnetoelectric (ME) multifferoics \cite{Tokura2006Science,Eerenstein2006Nature,Cheong2007NM,Fiebig2008JPD,Tokura2014RPP} brings up substantial questions of photo-induced dynamics due to more delicate interactions.  ME multiferroics of spiral-spin origin can lead to cross correlations between ferroelectric (FE) and spiral-spin orders, since the ionic displacement is a direct consequence of the inverse Dzyaloshinski-Moriya (DM) interaction \cite{KNB2005PRL,Mostovoy2006PRL}. Spin dynamics of ME multiferroics has been subject to focus \cite{Cheong2007NM,Pimenov2006NPhys,Senff2007PRL,Sushkov2007PRL,Takahashi2012NPhys} as it may become the platform for magnetoelectic memory \cite{Bibes2008NM}. Although insight into dynamical spiral-spin response has been given by using resonance THz or AC electric fields \cite{Pimenov2006NPhys,Senff2007PRL,Sushkov2007PRL,Takahashi2012NPhys,Hoffmann2011PRB,Lottermoser2009PRB}, a comprehensive understanding of ME dynamics is still lacking, primarily due to the lack of direct access to the FE order on various timescales \cite{Talbayev2015PRB,Johnson2015PRB}. However, their influence could be significant and unexpected states might be accessible from couplings between different degrees of freedom in non-equilibrium cases.  Therefore, we aim to unveil the photo-excited dynamics of spiral-spin induced ME coupling, spanning a wide range of timescale.

\begin{figure}[tb]
\begin{center}
\includegraphics[width=3.45in]{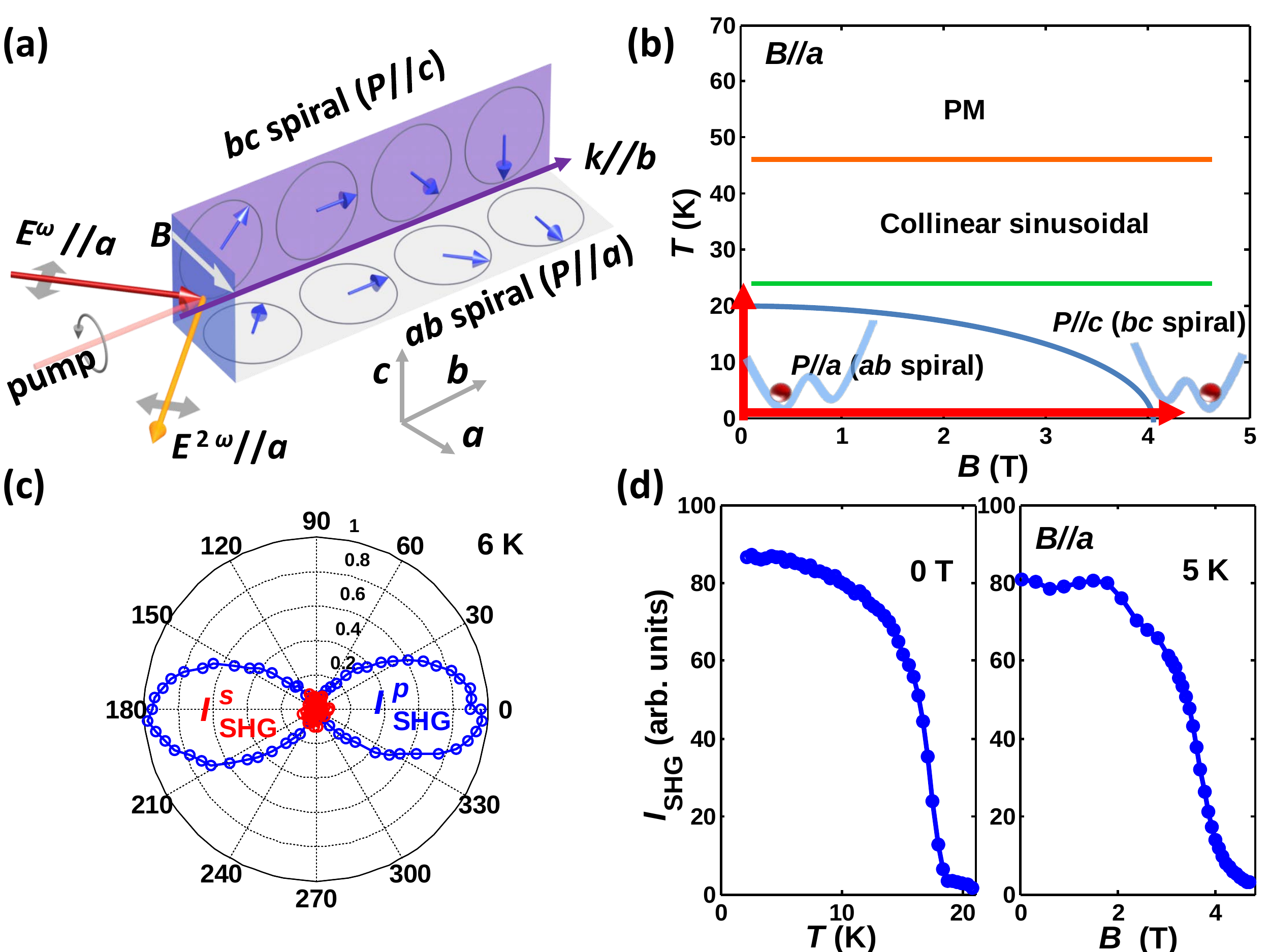}
\caption{ \label{fig1}(color online) (a) Schematic illustration for the time-resolved SHG (TR-SHG) in reflection configuration. Incident $p$ ($p$-in) and $s$ ($s$-in) photon polarizations correspond to $E^{\omega}$//$a$ and $E^{\omega}$//$c$.  The wavevector $k$ of the $ab$- and $bc$-spiral is along the $b$-axis. (b)Schematic phase diagram ($B$//$a$) of EYMO reproduced from Refs. \cite{Yamasaki2007PRB,Murakawa2008PRL}. The inset two free-energy landscapes illustrate the ground state of $ab$-(left) and $bc$-spiral (right). (c) Polar plot of the SHG (3.1 eV) intensity measured at 6 K, which detects the $ab$-spiral with $P$//$a$. $I_{\text{SHG}}^{p}$ and $I_{\text{SHG}}^{s}$ represent the $p$ ($p$-out) and $s$ ($s$-out) polarized SHG signal, respectively.  (d) Temperature ($T$) and magnetic-field ($B$//$a$-axis) dependence of the static SHG for the $p$-in and $p$-out configuration. The phase transitions from the $ab$- to the $bc$-spiral, observed in (d), are indicated as vertical (through temperature) and horizontal (through $B$) red arrows in the phase diagram of (b).     }
\end{center}
\end{figure}

Multiferroic perovskite manganites, $R$MnO$_{3}$ ($R$ being Tb, Dy, (Eu,Y), etc.), possess either of the $ab$- or $bc$-plane spin spiral states (abbreviated as $ab$-spiral and $bc$-spiral hereafter) with the propagation wavevector $k$ along the $b$-axis, (Fig. \ref{fig1}(a)). The two spiral states give rise to the FE polarization, $P$//$a$ or $P$//$c$ respectively, through the inverse DM interaction \cite{Mostovoy2006PRL,Sergienko2006PRB,KNB2005PRL}.  External magnetic field ($B$) can change the axis of spin cone and thus flip between the $ab$-spiral (under $B$//$c$) and the $bc$-spiral (under $B$//$a$) through a first-order phase transition \cite{Murakawa2008PRL}, as shown in Fig. \ref{fig1}(b).

To explore photoinduced phenomena of the two distinct ME states in perovskite manganites, we utilize the spin-induced FE polarization of Eu$_{0.55}$Y$_{0.45}$MnO$_{3}$ (EYMO) to generate second harmonic signal. Second harmonic generation (SHG) has been frequently applied to multiferroics of spiral-spin origin, such as MnWO$_{4}$ \cite{Meier2009PRL,Meier2009PRB,Meier2010PRB,Hoffmann2011PRB}, TbMnO$_{3}$ \cite{Matsubara2015Science}, TbMn$_{2}$O$_{5}$ \cite{Lottermoser2009PRB}, and CuO \cite{Hoffmann2012JPSP}. Ref. \cite{Fiebig2005JOSAB} and Ref. \cite{Denev2011SHGRev} have reviewed the techniques in depth. The time-resolved SHG (TR-SHG) can directly access the FE polarization (or equivalently the spiral-spin order here), spanning a wide range of timescale \cite{Sheu2013PRBR,Sheu2014Ncomms,Hoffmann2011PRB,Matsubara2009PRB}. Time-resolved research is practical for tracing dynamics of various order parameters \cite{Sheu2012APL,Wen2013PRL,Matsubara2009PRB,Ogawa2009PRB,Fiebig2008JPD,Ogawa2013APL,Sheu2014PRX,Li2013Nature}. In addition, the capability of generating highly non-equilibrated states with ultrashort pulses may lead to discovery of hidden states \cite{Ichikawa2011NM,Stojchevska2014Science,Oike2015PRB}, which would not be realized through conventional thermodynamic processes.

Single crystals EYMO of orthorhombic perovskite were grown by the floating zone method. The crystal of $ac$-surface was mechanically polished to $\sim$1 mm-thickness, and annealed at 750$^{\circ}$C in air for 12 hours to reduce the residual strain. It was mounted in a cryostat with a superconducting magnet and in contact with exchange He gas.

Our  TR-SHG experiments are based on an amplified Ti:sapphire laser system (1 kHz and $\sim$120 fs centered at 800 nm (1.55 eV)). Photon energies of 0.48-1.12 eV are generated by optical parametric amplifiers.  SHG signal is detected by a photomultiplier tube using lock-in techniques after filtering out the fundamental photon. Combination of optical chopper and mechanical shutter is employed for the long-time-delay measurements. All measurements are performed under zero field cooling.

\begin{figure}[tb]
\begin{center}
\includegraphics[width=3.45in]{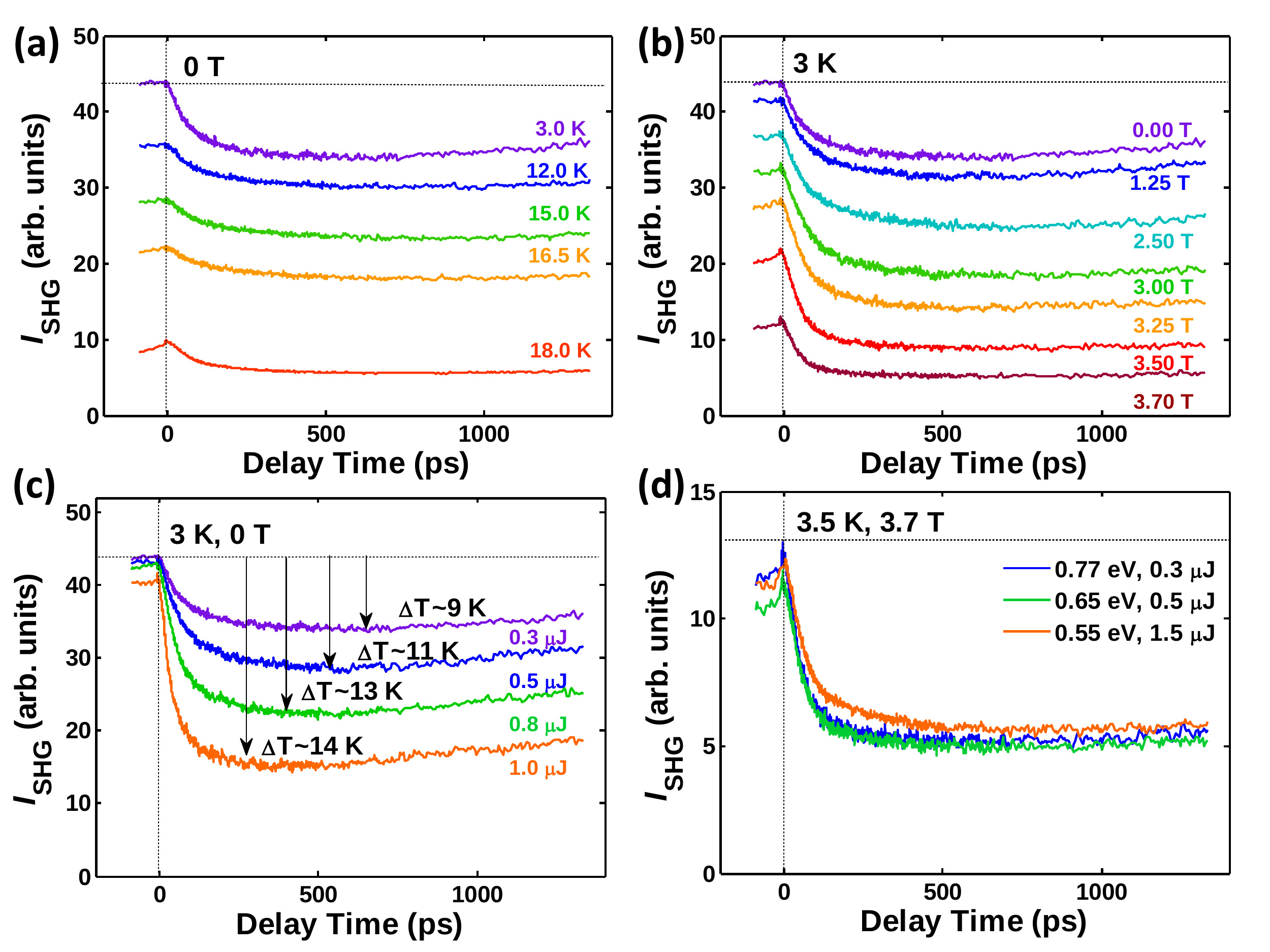}
\caption{ \label{fig2} (color online) Spin-lattice relaxation as the origin of TR-SHG response under low excitation ($<$1 $\mu$J). TR-SHG signal of the $p$-in (1.0 eV) and $p$-out (2.0 eV) setup (Fig. \ref{fig1}(a)), measured (a) at various temperatures in zero field,  (b) in various $B$ at 3 K, and (c) under different excitation intensity at 3 K in zero field. Photon energy of the pump is 0.77 eV. Here the 1-$\mu$J excitation corresponds to energy density 4.24 J/cm$^{3}$. (d) Traces of TR-SHG measured under different energies of pump photon. The excitation intensity was chosen so as to induce similar amount of reduction in $I_{\text{SHG}}$ to compare the dynamics. }
\end{center}
\end{figure}

At low temperature, EYMO has a ground state of the $ab$-spiral and $P$//$a$, of which the free-energy landscape is illustrated by the left inset of Fig. \ref{fig1}(b). As temperature rises,  the $bc$-spiral becomes thermodynamically more stable. The first-order phase transition between the two spin-spiral states occurs around 21 K, above which the FE polarization rotates to along the $c$-axis \cite{Murakawa2008PRL,Takahashi2009PRB}.  Figure \ref{fig1}(a) shows a schematic diagram of laser beam propagation with respect to the sample crystallographic axes. At 6 K without $B$, the polar plot of $p$- (//$a$) and $s$- (//$c$) polarized SHG is displayed in Fig.~\ref{fig1}(c). The polar patterns are consistent with the emergence of FE polarization along the $a$-axis.  More details of SHG tensor analysis are given in the supplemental material \cite{supplementary}. Figure \ref{fig1}(d) shows $p$-polarized SHG intensity, $I_{\text{SHG}}$, as a function of temperature (left panel) and $B$ applied along the $a$-axis (right panel), being consistent with occurrence of the $ab$-spiral in the phase diagram of Fig. \ref{fig1}(b). It is worthy noting that the applied $B$//$a$ rotates axis of spin cone and alters the polarization axis from $P$//$a$ to $P$//$c$, decreasing $I_{\text{SHG}}$ and changing the free-energy landscape toward the right inset of Fig. \ref{fig1}(b). The $B$-dependent SHG clearly identifies the $ab$-spiral through the induced FE polarization, and the associated SHG reduction under $B$//$a$ is a consequence of the $bc$-spiral formation due to the first-order phase transition. We note that the $bc$-spiral is not directly detectable by the SHG, likely due to the considerably small polarization. Discussion of missing SHG from $P$//$c$ is provided in the supplemental material \cite{supplementary}.

Upon relatively weak photoexcitation ($<$1 $\mu$J; the 1 $\mu$J excitation at 0.77 eV corresponds to energy density 4.24 J/cm$^{3}$), we observe a gradual reduction in $I_{\text{SHG}}$ as a function of delay time (Fig.~\ref{fig2}(a)).  The photoinduced depolarization completes in a time range of $\sim$50-100 ps, and the time constant is independent of temperature, magnetic field, excitation fluence, photon energy (Figs.~\ref{fig2}(a)-\ref{fig2}(d)) or light polarization (circular and linear; only circular is shown). The time constant for $I_{\text{SHG}}$ reduction is within the typical range of spin-lattice (S-L) relaxation in manganites \cite{Averitt2002JPCM,Ogasawara2003PRB,Muller2009NM}, in accord with the recent report \cite{Talbayev2015PRB} as well as our transient reflectivity (\dR~) data \cite{supplementary}. It is noted that the $I_{\text{SHG}}$ at negative delay time corresponds to the steady-state SHG at 1 ms after an 120-fs-pulse excitation and depends on temperature and magnetic field, just as in Fig.~\ref{fig1}(d).

Comparing our \dR~ with TR-SHG \cite{supplementary}, we can conclude that TR-SHG unambiguously probe spiral-spin dynamics without electronic contribution to the FE depolarization, due to the lack of fast electronic response $<$1 ps in TR-SHG while it is discerned in \dR.  In addition, the insensitivity of TR-SHG to pump photon energies again implies that the relaxation of spiral spin is not triggered by electronic contribution from specific optical transitions (including the $d$-$d$ transition here and the $p$-$d$ charge transfer in Ref. \cite{Talbayev2015PRB}). We can infer that the local spins (of $t_{2g}$ electrons) are still responsible for the FE polarization after electronic transition and/or transfer of excited $e_g$ electrons.  Thus the ferroelectric and spiral-spin order stay interlocked even in non-equilibrium case. Further insight into this is beyond the scope of the manuscript, and we provide the detail discussion and comparison in the supplemental material \cite{supplementary}.

The lack of electronic response and the characteristic timescale of S-L relaxation observed in TR-SHG imply that the depolarization is induced solely by the spin thermalization upon photoexcitation. The transient and quasi-equilibrium lattice temperature can be deduced from the TR-SHG traces at various temperatures (Fig.~\ref{fig2}(a)). At 3 K using 0.3 $\mu$J pulses with photon energy of 0.77 eV, the average change in temperature ($\Delta T$) is estimated to be 9 K from the reduced magnitude of $I_{\text{SHG}}$. At a fixed excitation fluence, $\Delta T$ reduces with increasing the sample base temperature (Fig.~\ref{fig2}(a)), whereas it stays almost constant while changing magnitude of $B$//$a$ (Fig.~\ref{fig2}(b)). This is reasonably anticipated from the increase of heat capacity at higher temperature. The time constant for the laser-induced heating process does not depend on the photon energy ranging from 0.55-1.55 eV (Fig.~\ref{fig2}(d)). Furthermore, the insignificant change in relaxation time near the transition, observed in both \dR~ and TR-SHG, indicates the first-order nature of phase transition, unlike the critical behavior of spin kinetics in manganites that possess second-order phase transitions \cite{Averitt2001PRL,Ogasawara2003PRB,Muller2009NM}.

\begin{figure}[tb]
\begin{center}
\includegraphics[width=3.2in]{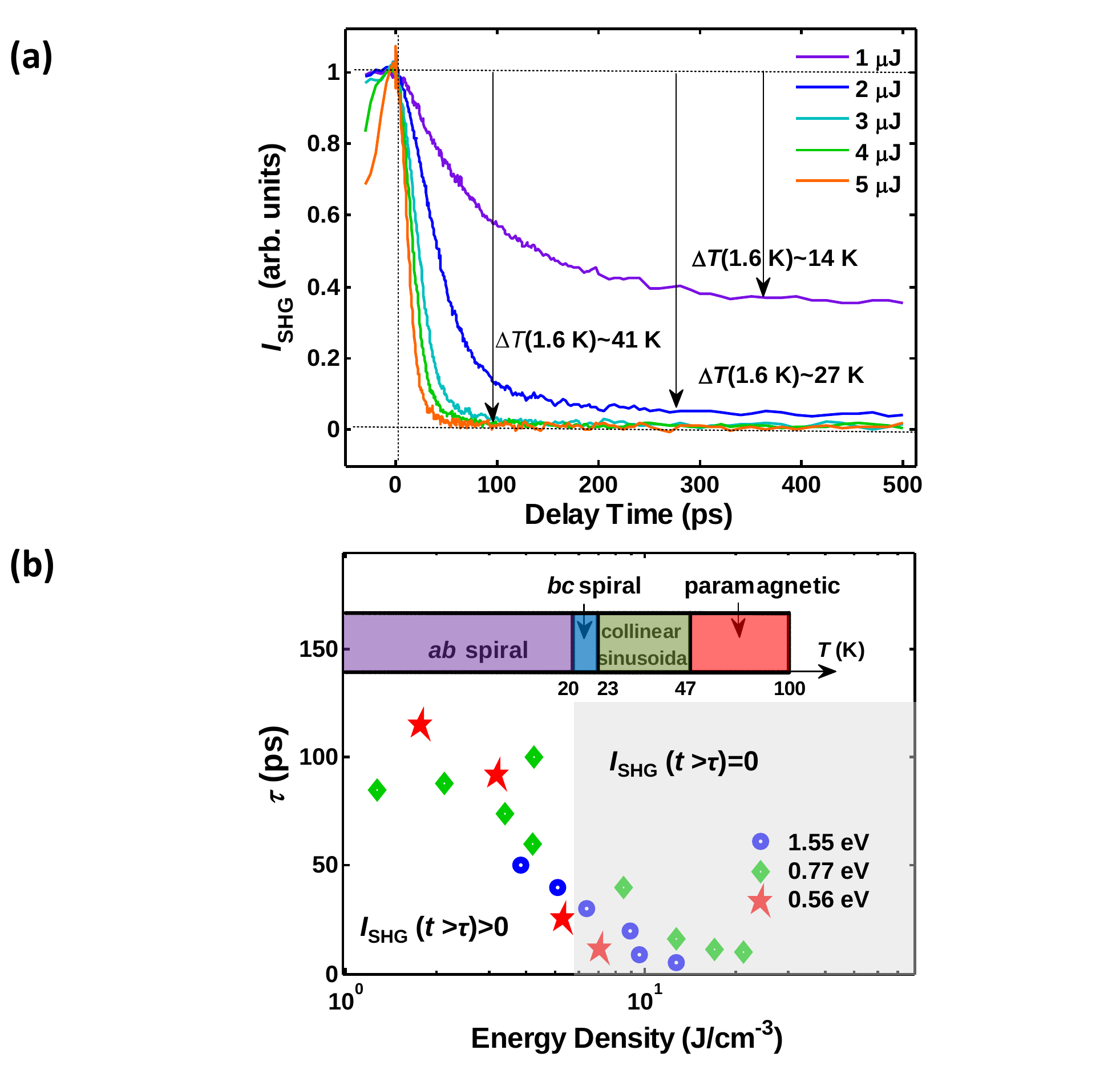}
\caption{\label{fig3}(color online) Formation of a metastable state under high excitation.  (a) TR-SHG of $p$-in (1.0 eV) and $p$-out (2.0 eV) setup measured at 1.6 K under various pump intensities (above 1 $\mu$J). The vertical arrows indicate the estimated change in lattice temperature. (b) Time constant of FE depolarization versus excitation energy density at three photon energies. The time constant is extracted from the fit of single exponential decay. The shaded grey area covers the excitation density leading to the complete reduction of the SHG (i.e. phase transition to the $bc$-spiral and/or higher-temperature phases).  The inset color blocks represent the corresponding lattice temperatures and the respective multifferroic/magnetic phases therein.
}
\end{center}
\end{figure}

\begin{figure}[tb]
\begin{center}
\includegraphics[width=3.2in]{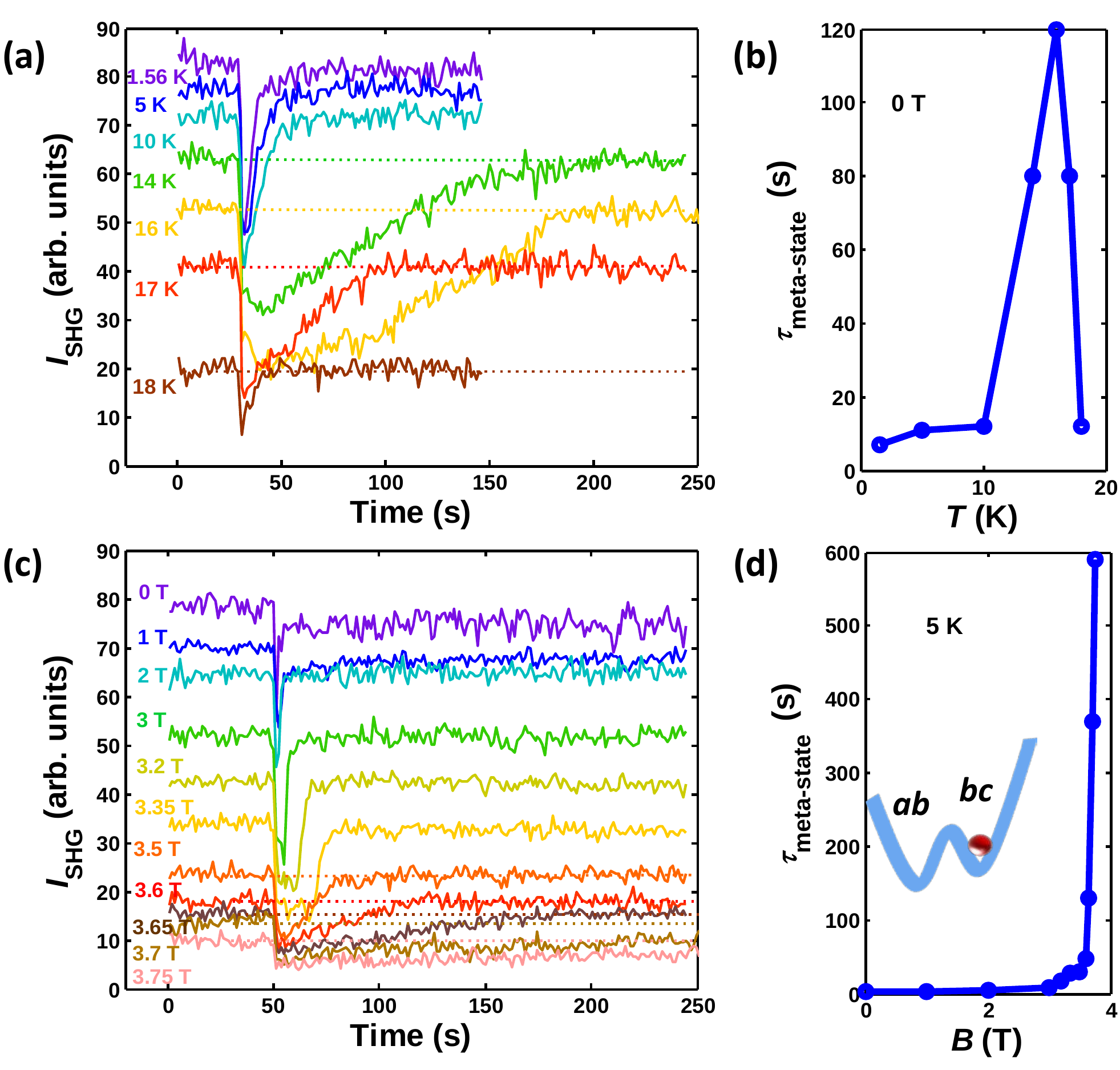}
\caption{\label{fig4}(color online) Photo-creating the metastable $bc$-spiral from the ground state of $ab$-spiral.  The high excitation in the grey area of Fig. \ref{fig3}(b) creates a metastable spin state, which is probed by the long-lived reduction in spin-induced SHG: (a) TR-SHG measured at various temperatures and  (b) the corresponding lifetimes of metastable states. (c) The recovery of photo-created metastable states under application of various  $B$//$a$ at 5 K, and (d) the corresponding lifetimes. The inset illustrates a schematic free-energy landscape against the canting angle of the spin-spiral plane around the $b$-axis and a formation of the metastable $bc$-spiral.
}
\end{center}
\end{figure}

Intriguing phenomena are uncovered as the excitation density increases further: photo-depolarization of the spin-induced FE becomes faster and finally the $I_{\text{SHG}}$ disappears completely, as seen in Fig.~\ref{fig3}(a). To discern the origin of this feature, we plot the decay time constant as a function of excitation energy density in Fig. \ref{fig3}(b), which reveals a threshold behavior irrespective of pump photon energies. Above the threshold the TR-SHG goes to zero during a short time ($<$20 ps) and remains at zero for long time ($>$50 ps), whereas below the threshold TR-SHG never reduces to zero completely, and slow S-L relaxation dominates the FE depolarization, irrespective of sample temperature and magnetic field. Thus such a threshold energy-density differentiates the "low" and "high" excitation regime, as illustrated in the white and grey area of Fig.~\ref{fig3}(b). The correlation with pump energy density as well as the irrelevance of pump photon energy makes the TR-SHG an ideal thermometer of nonequilibrium spin subsystem in low-excitation case, allowing us to determine the transient spin temperature. In high-excitation case, we can estimate lattice temperature (inset of Fig.~\ref{fig3}(b)) by linear extrapolation since the $I_{\text{SHG}}(t)$ disappears completely and no longer works as the spin thermometer. We find that once the transient temperature ($T+\Delta T$) is above $T_{\text{c}}$ of the $ab$-spiral, the $I_{\text{SHG}}$ decreases faster and then disappears completely for a prolonged time ($>$50 ps).

As the pump intensity approaches the threshold, a subtle behavior occurs before the zero delay time (Fig.~\ref{fig2}(c)); we notice a small reduction of $I_{\text{SHG}}$, which becomes significant in the high excitation region (Fig.~\ref{fig3}(a)), e.g. $I_{\text{SHG}}^{1\mu\text{J}}(t<0)<I_{\text{SHG}}^{0.3\mu\text{J}}(t<0)$. This indicates that the $ab$-spiral does not recover within 1 ms (at 1 KHz laser repetition), which signals the situation generally not explored in the time-resolved optical experiments.  The observed reduction is not simply attributable to accumulated lattice thermalization, rather pointing to the emergence of metastable spin orders, as discussed below.

To investigate the subtle SHG reduction before the zero time delay (i.e. existing at 1 ms after photo-excitation), we reduced the pump repetition rate to allow the system to recover the thermodynamical ground state ($ab$-spiral) at the base temperature. We also adjust the pump intensity around 24-28 $\mu$J to unveil dynamics of the metastable state. The resultant SHG traces at various temperatures (Fig.~\ref{fig4}(a)) and under different $B$//$a$ (Fig.~\ref{fig4}(b)) reveal dramatic change in the recovery time, spanning from a few seconds up to several minutes. As the base temperature approaches the phase transition, the recovery time becomes longer, as shown in Fig. \ref{fig4}(a). Similarly, the application of $B$//$a$ lengthens the recovery time (Fig.~\ref{fig4}(b)). We thus exclude the lattice residual heating as the origin of SHG reduction, since it would not result in long recovery time with such strong dependence on base temperature of crystal nor on applied magnetic field. Since the external $B//a$ energetically favors the $bc$-spiral and the (thermal) recovery dynamics inevitably goes through the phase of $bc$-spiral, the prolonged response after an intense pump excitation likely involves the transition of the photo-generated $bc$-spiral to the original $ab$-spiral.

The remarkably long timescale observed here indicates formation of a metastable state, reachable by supercooling the system through a first-order phase transition, from which a system cannot escape at low enough temperatures due to the high free-energy potential barrier \cite{Barrat1996JPA,Berthier2011Review}. In other words, the supercooled state preserves partially or dominantly the order of the high-temperature state, because the timescale to form the critical size of the thermodynamical ground state becomes long enough. We can estimate the cooling rate from the trace of 1-$\mu$J excitation shown in Fig.~\ref{fig2}(c). The residual heating of $\sim$4 K is estimated by the SHG signal just before the time zero, while the transiently elevated temperature $\Delta T$($t>$100 ps) is $\sim$14 K; this implies that the cooling rate is as high as 10$^{4}$ K/s at 1 kHz pump repetition. The supercooling effect is often observed in temperature- and/or magnetic-field-hysteresis features of other multifferoics of $R$MnO$_{3}$ \cite{Kimura2005PRB} and charge-ordered manganites with first-order phase transitions \cite{Tokura2006RPP}.

We can use this estimate to differentiate the two processes of SHG reduction; i.e. lattice heating versus a formation of the metastable state. Using linear extrapolation of the relation between light absorption and heat for the present experimental condition in Fig.~\ref{fig4}, we estimate the lattice heating up to $\sim$ 300 K which cools to the base temperature in $\sim$ 30 ms after photoexcitation. The recovery time, shown in Figs.~\ref{fig4}(b) and \ref{fig4}(d), is near four orders of magnitude longer than this lattice cooling time; thus the magnetically ordered phase (e.g. spin-spiral) should be restored.  We do not expect that a change in the recovery time occurs in heat dissipation owing to variation in diffusion. If it were the case, we would observe a gigantic critical-slowing-down during S-L relaxation in low-excitation case, Figs.~\ref{fig2}(a) and \ref{fig2}(c). Instead, the prolonged SHG recovery can be explained in terms of the trap of $bc$-spiral in free-energy potential well (see the schematic inset of Fig.~\ref{fig4}(d)), frozen by supercooling. Compared with typical timescale of tens milliseconds for nucleation and domain growth upon fast electrical poling to induce spiral-spin flip \cite{Hoffmann2011PRB,Baum2014PRB}, cooling from the upper temperature of $bc$-spiral to the temperature possessing the thermodynamical ground state of $ab$-spiral takes less than 1 ms, which is too short to form large size of $ab$-spiral domain. As for the origin of this minima of free-energy potential well, it could be the high temperature collinear spin state or some other states. However, during supercooling across the $T_{\text{c}}$ of $bc$-spiral, the application of $B$//$a$ can energetically favor the $bc$-spiral to be trapped momentarily  as the transient state. Further cooling at high speed can hardly overcome the potential barrier between $ab$- and $bc$-spiral, and then the field activated stable phase of the $bc$-spiral cools down and is fixed by the barrier, forming a metastable state at the base temperature (inset of Fig.~\ref{fig4}(d)). Therefore, when the temperature is near $T_{\text{c}}$ or $B$//$a$, the $bc$-spiral is dominant more than other collinear sinusoidal or disorder states during supercooling.

In conclusion, by probing the temporal variation of second harmonic generation (SHG) from the spiral-spin-induced electric polarization,  we have investigated dynamics of photoexcited multiferroic state in \EYMO~ for a wide range of timescale from 10$^{-12}$ to 10$^{3}$ seconds. We confirm that there is no specific electronic contribution to FE depolarization apart from the thermalization due to spin-lattice (S-L) relaxation on all timescales. We find that a metastable $bc$-spiral ($P$//$c$) can be photo-created from the $ab$-spiral ($P$//$a$) through supercooling from the photo-thermalized state. We observe the two distinct dynamics differentiated by a threshold energy density of photo-excitation: Below the threshold the spiral spin relaxes through S-L coupling and the whole system then cools back to the original temperature while preserving $P$//$a$. Above the threshold, by contrast, we observe a remarkable slow SHG recovery with strong dependence on temperature and magnetic field applied along the $a$-axis. This feature together with the estimate of cooling rate allows us to conclude that supercooling a multiferroic manganite using femtosecond pulses creates a metastable $bc$-spiral state with $P$//$c$. Our study thus provides an insight into the switch of multiferroics of spiral-spin origin and will pave a new avenue toward nonvolatile memory-storage functionality \cite{Oike2015PRB}.

\acknowledgments{The authors thank F. Kagawa for the stimulating discussion and M. Ishida for the technical help. This research is granted by the JSPS Grant-in-Aid for Scientific Research(S) No. 24224009. Y.M. Sheu acknowledge the support from the Taiwan Ministry of Science and Technology, grant No. MOST 104-2112-M-009-023-MY3.   }

%\bibliography{EYMO}
%

\end{document}